# Ferroelectric dynamic-field-driven nucleation and growth model for predictive materials-to-circuit co-design

Yi Liang[1, 2,*], Soohyeon Kim[3], Tony Chiang[1, 2], Megan K. Lenox[4], Ian Mercer[5], John J. Plombon[6], Jon-Paul Maria[5], Jon F. Ihlefeld[4, 7], Wenhao Sun[1], Wei Lu[3], John T. Heron[1, 2,*]

[1]Department of Materials Science and Engineering,
University of Michigan, Ann Arbor, MI 48109, USA

[2]The Ferroelectronics Laboratory, University of Michigan, Ann Arbor, MI 48109, USA

[3]Department of Electrical Engineering and Computer Science,
University of Michigan, Ann Arbor, MI 48109, USA

[4]Department of Materials Science and Engineering,
University of Virginia, Charlottesville, VA 22904, USA

[5]Department of Materials Science and Engineering,
The Pennsylvania State University, University Park, PA, 16802 USA

[6]Technology Research, Intel Corporation, Hillsboro, OR 97124, USA

[7]Charles L. Brown Department of Electrical and Computer Engineering,
University of Virginia, Charlottesville, VA 22904, USA

Corresponding authors: liangyy@umich.edu jtheron@umich.edu

**Abstract:** Real ferroelectric devices operate under mixed and distorted time-varying voltages, yet the standard nucleation-growth frameworks used to interpret ferroelectric switching — most notably the Kolmogorov-Avrami-Ishibashi (KAI) and nucleation-limited switching models (NLS) — are derived under the critically limiting assumption of a constant electric field. Thus, the prevailing interpretation of ferroelectric switching dynamics fails under real operating conditions. Here we introduce a compact dynamic-field-driven nucleation and growth (DFNG) model that enables quantitative fits to switching transients across multiple ferroelectric materials to extract time-varying domain wall velocity and growth dimensionality, even under arbitrary voltage waveform. This capability then motivates its use in device modeling under complex signals spanning disparate time and frequency scales. Coupling the compact model to application-related waveforms and circuit-level simulation platform facilitates a predictive materials-circuit co-design framework by linking nucleation and growth parameters to memory window, disturb error, speed, and energy dissipation for next-generation ferroelectric technologies.

## 1. Introduction

The low-power, high-speed, and non-volatile characteristics of ferroelectric materials have enabled device concepts[1–3] that can surpass conventional CMOS technologies[4–6]. Key to the ferroelectric computing technologies is the manipulation of polarization under ultrafast mixed-frequency and time-dependent voltage waveforms (**Figure 1a**). The time evolution of polarization transient composes the intricate metrics of ferroelectric materials and devices, such as latency, energy, disturb errors, and readout signals. To assist high throughput material design and to optimize device operation protocols, a compact model is needed to predict physical response of ferroelectrics under the dynamic electric fields.

Existing models, however, are unable to describe the nucleation and growth processes which drive ferroelectric switching under a time-varying voltage, largely because ferroelectric polarization evolution is hysteretic and depends on the full history of applied electrical stimuli. For example, the seminal Kolmogorov-Avrami-Ishibashi (KAI) model[7–9] provides an explicit description of the polarization transient under a constant applied voltage. Even when a voltage dependence is inappropriately forced, the polarization is only dependent on instantaneous voltage value and ignores the prior voltage path (**Figure 1b** and **Supplementary Note 1**). This deficiency is also found in the nucleation-limited switching (NLS) model[10], which simulates a switching event from a distribution of local nucleation times under a constant applied field and neglects domain growth. In addition, the distribution function in the NLS model is ad hoc and thus usually leads to overfitting of experimental transients, thereby overlooking physical rationalization. On the other hand, time-dependent Ginzburg-Landau-Devonshire[11–13], Preisach, or other equivalent-circuit models, can be numerically driven by any waveform, but they neglect the necessary nucleation and growth physics or are generally too complex to serve as compact, closed-form fitting tools for experimental transients.

In this work, we present a dynamic-field-driven nucleation and growth (DFNG) model that is universally applicable to distinct materials and arbitrary waveforms (**Figure 1b**). We first demonstrate the capability by self-consistently fitting experimental polarization switching transients for polycrystalline $Hf_{0.5}Zr_{0.5}O_2$ (HZO), single-crystalline $BaTiO_3$ (BTO), and $Al_{0.92}B_{0.08}N$ (AlBN) thin films with as few as 3 parameters. The model robustly extracts intrinsic materials parameters (growth dimension, nucleation density, and Merz activation field), even when there are mixed time-scale changes in the voltage profile due to the rise time of electronics (10-100 ps) and slower circuit parasitic effects (1-10 ns), making materials-level inference possible even under practical measurements with non-idealities. We then derive domain wall velocity and domain size of HZO, which motivates a speculation of domain-growth-limited switching mechanism with instantaneous nucleation corroborated by experimental findings, contrary to the proliferated belief of nucleation-limited switching. Furthermore, the model generates realistic material response to time-varying mixed-frequency waveforms which can readily incorporate with circuit-level modeling to optimize circuit design and operation protocols. Finally, the model is used to map material parameters to hysteresis loops and technology metrics, providing insight on design principles based on application requirements. We further demonstrate the capability of large-scale circuit simulation by implementing the DFNG model and including other circuit elements with SPICE simulator. Together, our study for the first time establishes a universal, physics-grounded framework that links nucleation and growth dynamics to device-level performance metrics, offering a foundation for elucidating material fundamentals and a rational co-design platform for next-generation ferroelectric technologies.

## 2. Results and Discussion

### 2.1 A compact yet generalized nucleation and growth model

Our DFNG model extends the traditional Avrami theory of nucleation and growth to capture the ultrafast voltage profiles driving a ferroelectric phase transformation by including two important physical considerations: 1) Nucleation is either instant or adopts a changing steady-state nucleation rate governed by thermodynamic barrier that is dependent on the time-varying voltage profile; 2) a non-linear voltage-dependent growth rate based on Merz law[14].

We build our derivation from Cahn's time-cone model[15], which relates the transformed fraction, $f(t)$, of a system undergoing a phase transition to the number of nuclei, $\langle N(t)\rangle$, in a cone of space-time that precedes the time and point of space where nucleation may occur (reverse time cone) by

$$f(t) = 1 - \exp[-\langle N(t)\rangle] \tag{1}$$

Where,

$$\langle N(t)\rangle = \gamma \int_0^t J(\tau) \left[ \int_\tau^t v(t')dt' \right]^d d\tau \tag{2}$$

Here $d$ is domain growth dimension, and $\gamma$ is the geometric factor of domain growth, e.g. $\gamma = 1$ for 1-dimensional, $\pi$ for 2-dimensional, and $\frac{4}{3}\pi$ for 3-dimensional. $\gamma$ and $d$ can be applied to mixed dimensions. $v(t')$ is the domain wall velocity at time $t'$. We introduce Merz law[14] to describe the velocity, namely $v(t') = v_0 \exp\left[\frac{-V_a}{V(t')}\right]$, where $v_0$ is the domain wall velocity at infinite voltage approximated as the speed of sound, $V(t')$ is the instantaneous voltage and $V_a$ is the activation voltage, corresponding to an average potential that the domain wall depins from a defect site. $J(\tau)$ is the nucleation rate at time $\tau$, which is then discussed in two cases, instantaneous nucleation and homogeneous nucleation.

For instantaneous nucleation, $J_{Inst}(\tau) = \rho_1 \delta(0)$. Substituting into **Equation 2**, $\langle N(t) \rangle$ is acquired.

$$\langle N(t) \rangle = A^d \left[ \int_0^t \exp\left[ -\frac{V_a}{V(t')} \right] dt' \right]^d , \tag{3}$$

Here, $A^d = \gamma \rho_1 v_0^d$, and $\rho_1$ is the density of pre-existing nuclei. Thus, parameter $A$ is an indicator of instantaneous nucleation density.

For the homogeneous nucleation, we adopt an expression from the Janovec-Kay-Dunn (JKD) model[16,17] to calculate the nucleation rate. In the JKD model, the probability of nucleation per site is given by $\exp\left(-\frac{\Delta G^*}{k_B T}\right)$, where $\Delta G^*$ is the critical Gibbs free energy to create a nucleus. Assuming a cylindrical nucleus that extends across the thickness and ignoring the depolarization field, $\Delta G^*$ is given by $\frac{\pi t_{FE}^2 \sigma^2}{2 P_S V(\tau)}$. $t_{FE}$ is the thickness of the ferroelectric layer, $\sigma$ is the domain wall energy, $P_S$ is the spontaneous polarization, and $V(\tau)$ is the instantaneous voltage when nucleation happens at time $\tau$. Another physical process involved in homogeneous nucleation is the attachment of dipoles, the rate of which is calculated by the product of dipole flipping frequency $\omega_0$ and the probability of a successful flip given by $\exp\left(-\frac{V_a}{V(\tau)}\right)$. Then the homogeneous nucleation rate at time $\tau$ is the product of site density, nucleation probability and dipole attachment rate[18],

$$J_{Homo}(\tau) = \rho_2 \omega_0 \exp\left[ -\left( \frac{\pi t_{FE}^2 \sigma^2}{2 P_S V(\tau) k_B T} + \frac{V_a}{V(\tau)} \right) \right], \tag{4}$$

where $\rho_2$ is the homogeneous nucleation density. At constant temperature, the expression of $\langle N(t) \rangle$ becomes:

$$\langle N(t) \rangle = B^d \int_0^t \exp\left[ -\frac{V_a + \sigma'^2}{V(\tau)} \right] \left[ \int_\tau^t \exp\left[ -\frac{V_a}{V(t')} \right] dt' \right]^d d\tau , \tag{5}$$

where $B^d = \gamma \rho \omega_0 v_0^d$, $\sigma'^2 = \frac{\pi t_{FE}^2 \sigma^2}{2 P_S k_B T}$. We note that the integral forms in **Equation 3** and **5** preserve voltage path dependent switching transients.

### 2.2 Material parameters from measurements in dynamic field across materials

We begin by applying the model to the HZO system. We experimentally measure ferroelectric switching transients in 3-μm diameter capacitors by simultaneously acquiring the voltage across the ferroelectric capacitor ($V_{FE}$), the ferroelectric switching current *df/dt* (normalized by $2P_S$), and the polarization (transformed fraction *f* when normalized by $2P_S$) as a function of time (Figure 1a and **Figure 2a**). Details of the transient measurements are given in the Methods section and elsewhere[19]. We find that the model best fits the experimental data when an instantaneous nucleation rate is used (**Figure S1**), indicating the presence of preexisting nuclei. These nuclei are possibly permanent dipoles formed by bound charges at the electrode-insulator interface, grain boundaries and phase boundaries (hence heterogeneous). The fit yields $d = 0.699 \pm 0.003$, $V_a = 12.75 \pm 0.09$ V, $A = 3236 \pm 191$ ns$^{-1}$. These parameters are extracted via a global fit across four different supply voltage profiles while feeding the measured real-time voltage as an input. This ensures the extracted parameters are intrinsic to the material, independent of the variance in applied voltage. The values of these parameters reflect their physical origins and will be discussed in a few paragraphs.

The expression of Equation 3 for HZO is compact, as it only involves three parameters $A$, $V_a$, and $d$, yet able to capture the full experimental switching transient particularly through the rapid voltage variations that occur at sub-ns and ns scales (Figure 2a). In real circuits, the rise of the voltage pulse is skewed by the finite rise time of the pulse generator and the dielectric current, which here is ~ 200 ps. Interestingly, ferroelectric switching occurs during this rapid voltage rise and produces a dynamic Avrami exponent (exceeding 4) at the beginning. The DFNG model captures this behavior and outputs a large initial Avrami exponent because the domain wall velocity accelerates non-linearly upon the rapid rise of the voltage. After, the Avrami exponent relaxes back to an interpretable growth dimension value as the voltage settles. This result indicates that apparent super-physical Avrami exponents frequently reported in the literature naturally arise from dynamic-field effects. After the initial voltage rise, significant ferroelectric switching current is generated and flows through the series resistor in the circuit for larger voltage amplitudes, leading to a sluggish rise to the set voltage (between the green and grey dashed lines in Figure 2a). This distortion of the voltage waveform spans through almost the entire switching process, which highlights the necessity of incorporating a time-dependent voltage for modeling ferroelectric switching transients. The deviation from the ideal square pulse also implies the common misuse of the conventional KAI and NLS model where constant voltage is assumed. Contrarily, our model allows for robust physical parameter extraction even when there are ultrafast changes in the voltage profile due to the rise time of the electronics (10-100 ps scale) and slower circuit parasitic effects (1-10 ns scale) making possible materials-level inference under unideal measurement conditions.

To demonstrate universality, we apply the model to single-crystalline BTO with both heterogeneous and homogeneous nucleation included, by summing Equation 3 and 5, and involving 5 fit parameters. The model fits to the whole transient of a 3-μm BTO capacitor despite rapid voltage variations and for all voltage profiles (**Figure 2b**). A low $V_a$ of 0.536 V, sparse heterogeneous nucleation density of ~ $10^{-4}$ $nm^{-d}$, and domain wall energy of 4.4 $mJ/m^{-2}$ are determined and consistent with the single-crystalline nature of BTO (**Supplementary Note 2**). Additionally, the model is applied to an ultrathin AlBN capacitor (10 nm thick, 100 μm in diameter) as a representative wurtzite ferroelectric system (**Figure 2c** and **Supplementary Note 3, Figure S2 and S3**). The fitting results indicate that in this thickness regime, AlBN switches predominantly via a heterogeneous nucleation mechanism with a high activation voltage (66.0 V). Although heterogeneous nucleation typically implies defect-assisted switching, the defects that can initiate nucleation are rare (~$10^{-5}$ $nm^{-d}$). The combination of low nucleation density and high activation voltage accounts for the large switching field (7-9 MV/cm) and long switching times (10s μs), in good agreement with reported literature values[20–23].

### 2.3 Uncovering microscopic switching mechanisms from model analysis

Nucleation sites, domain wall velocity, and domain sizes that can be visualized or inferred with modern techniques such as piezoresponse force microscopy (PFM) and advanced transmission electron microscopy[24–27], serve as observable quantities to verify physics postulated in theoretical models. With the DFNG model, we derive physical values for these quantities and further posit a growth-limited switching mechanism, which can be corroborated by domain imaging techniques and simulations.

By assuming a $v_0$ value of 6000 m/s[28], domain wall velocity $v$, nucleation rate $J$, and domain radius $r$ can be directly output from the model (**Figure 3**). The domain wall velocity of HZO (**Figure 3b**) during switching increases from 0 to 1.2 m/s (low voltage, light purple) and 14.4 m/s (high voltage, dark purple) resulting from the high activation field for depinning in a defective medium. Despite the slow growth rate, HZO exhibits a comparable switching time to BTO because of a much higher nucleation density (**Figure 3c**). Heterogeneous nucleation density of HZO is estimated to be ~1 $nm^{-d}$. Note that the heterogeneous nucleation density is the volumetric number of nucleation sites averaged over the device area. Therefore, multiple nucleation defects can exist in one grain as they reside at the grain boundaries, intragranular phase boundaries, and the ferroelectric-electrode interface. The domain wall velocity also leads to distinct profiles in time cones and domain sizes (**Figure 3d**). In contrast to the straight time cone profile in the conventional KAI model, real materials possess a curved lateral surface due to a time-varying domain wall velocity. Domain radius $r$ of HZO is estimated to be several nanometers. The small domain size of HZO provides an effective alternative to the NLS model's claim that domain wall motion is confined within a grain. Our model posits that nucleation is instantaneous and domain walls are mobile across grain boundaries albeit encountering a high Merz barrier. Therefore, a switching paradigm that is limited by domain growth (Figure 3d, top right) is proposed in contrast to the NLS picture (Figure 3d, bottom right) where domain wall mobility is bounded by polycrystallinity and granularity. Our picture is commensurate with a recent electron holography experiment[25], which shows a domain size of a few nanometers and the lateral growth of domains to the neighboring grain, validating the physical foundation of our model. Additionally, multiple PFM studies also emphasized the non-negligible role of lateral domain growth in the switching in $HfO_2$-based ferroelectrics[26,29,30]. On the other hand, complementary experimental and computational works report a transition between KAI-type and NLS-type switching kinetics via thickness scaling, crystallinity tuning and temperature control[22,29,31,32] within one material system, indicating a competition in the dominance of nucleation and growth, as well as a potential competition between different nucleation mechanisms. These findings motivate a physics-based analytical model to cover a continuum of switching kinetics and extract the key parameters governing the interplay between nucleation and growth, rather than relying solely on phenomenological descriptions tied to a single limiting regime.

### 2.4 Polarization behavior under realistic complex circuit voltages

Nowadays, ferroelectric materials have attracted significant interest for neuromorphic computing as a compute-in-memory (CiM) technique to overcome the von Neumann bottleneck, as the stored polarization charge naturally encodes synaptic weights. FeFET (ferroelectric field effect transistor), FeCAP (ferroelectric capacitor), and FTJ (ferroelectric tunnel junction) crossbar arrays are developed for non-destructive readout of the states[33–35]. The operation of these devices requires deliberately designed mixed signal waveforms to achieve correct and efficient functionality[36]. The previous void of a compact and physical model of polarization switching under dynamic electric field hinders the incorporation of ferroelectric components into circuit-level modeling and thus optimization. Here we show that the DFNG model can produce realistic material responses to a potential complex dynamic waveform.

We use the mixed signal nature of a small signal capacitance-voltage $C(V)$ hysteresis loop measurements as an example (**Figure 4**), where a low-amplitude AC signal superimposed on a DC bias is used. This measurement well represents the complexity of the waveform used in real-world circuits and the small-signal capacitance is recently proposed as an efficient readout mechanism for charge-based neuromorphic computing[37,38]. In **Figure 4a**, the model produces the ideal hysteresis, where leakage and internal fields are ignored. It replicates the empirical trend that increased AC frequency leads to a higher coercive field and lower peak capacitance defined by the $C(V)$ loops. Moreover, the polarization ($f$) and current ($df/dt$) traces in **Figure 4b** essentially indicate the accumulated charges and current flow which are potential readout signals in a circuit with a ferroelectric element. We note that the polarization and current drift during the AC oscillation causes a time-dependent variation in the readout capacitance values and may compromise the read margin if multi-level operation is performed on the FeCAP. The variance of capacitance is most significant around the DC voltage where polarization switching occurs most rapidly.

**2.5 Materials-to-circuit co-design platform enabled by the model**

We next present a potential routine for evaluating material eligibility for technologies and accelerating synthesis optimization. From a ferroelectric material synthesis prospective, polarization-voltage $P(V)$ hysteresis measurement is the most typical diagnosis. We therefore vary the material parameters and map the corresponding $P(V)$ loops in **Figure 5a** (**Supplementary Note 4** and **Figure S4**, **Figure S5**). It qualitatively shows that decreasing heterogeneous nucleation sites or increasing depinning barrier leads to wider loops, and that decreasing growth dimension leads to skewed loops. Meanwhile, pulse operation (**Figure S6**) and high frequency ramps, as frequently used waveforms in applications, are performed to extract metrics of interest. We then grade the material parameters following the procedure below (and **Supplementary Note 5**):

1) Determinism under CMOS conditions: 99% switching fraction should be completed within 10 ns (delay) with a pulse amplitude ($V_{min}$) $\leq$ 1.5 V[4].
2) Memory stability against disturb: characterized by the polarization loss fraction $\Delta f$ under a write disturb pulse in the $V_{DD}/2$ scheme[39]. Nominally $\Delta f \leq 5\%$ is acceptable for memory device such as FeRAM (ferroelectric random-access memory), and $\Delta f \leq 1\%$ for CiM devices as they encounter more frequent write operations.
3) Energy and delay for CiM under write operation.
4) Memory window for FeCAPs[40]: defined by normalized peak differential capacitance, $C_{norm} = \left(\frac{df}{dV}\right)_{f=0.5} = \frac{1}{P_S t_{FE}} \epsilon_{FE}$. $\epsilon_{FE}$ is the dielectric constant of the ferroelectric (subtracting the linear part).
5) Read margin / Memory window for FeFETs[41]: defined by the coercive voltage, $V_C = V_{f=0.5}$.

We locate the loop of the current HZO sample according to its fit parameters (Figure 5a, purple star). This immediately suggests that a modest increase in nucleation density and growth dimension would adapt the material for a CiM application (orange star). Physically, such tuning could be achieved by introducing small dipole clusters and removing some grain boundaries, as charged defects act as heterogeneous nucleation sites while mesoscopic grain boundaries truncate the time cone and reduce growth dimension. The corresponding ranges of energy and delay are indicated by the red dots beside the loop. The orange star matches the loop in Figure 5a to a FeFET application with corresponding metrics in **Figure 5b**. Furthermore, Figure 5b provides the tunable range and pairwise correlation of the metrics which allow inverse design strategy and fast estimation of performance. For example, FeFETs favor materials with large $V_C$, outlined by the orange box. These selected sets are bound to good control of memory stability and tight distribution of energy dissipation, leaving delay the only variable of concern. Similarly, FeCAP devices favor high capacitances outlined by the green box. However, achieving these capacitance values requires high growth dimension ($d$ > 1.6) and low Merz barrier ( $V_a$< 4 V), which can be extremely demanding in polycrystalline materials. More discussion is attached in **Supplementary Note 5, Table S1 and S2**. Other general trends can also be concluded. The correlation between $V_{min}$, delay, and energy indicates that material parameters enabling lower operating voltages tend to reduce the switching energy, as the switching time decreases exponentially. The relation between $\Delta f$ and other metrics is less defined as it is subject to a different pulse amplitude, suggesting that disturb immunity cannot be inferred from other measurements and requires a dedicated examination.

To facilitate a materials-circuit co-design framework and enable large scale circuit-level modeling, we further establish a SPICE model that incorporates the dynamic ferroelectric elements described by the DFNG model (**Figure 6a**). On a single device level, the equivalent FeCAP sub-circuit consists of an intrinsic linear capacitor in parallel with a voltage-controlled current source representing the ferroelectric elements. Details of the SPICE FeCAP cell model can be found in **Supplementary Note 6 and Figure S7**.

As shown in **Figure 6b**, the dynamic switching current as well as the polarization state produced by the SPICE FeCAP model accurately match the experimentally measured results under similar waveforms, demonstrating the capability of transitioning the DFNG model to SPICE device models for practical circuit simulations. Building upon this single-device foundation, the SPICE simulation framework is expanded to evaluate the FeCAP crossbar array architecture that takes into account the dynamic ferroelectric switching behaviors and parasitic effects in the crossbars (**Figure 6c**, left). The simulation produces a switching delay of 15.6 ns on a 3×3 array, and 15.7 ns on an expanded 32×32 array, due to RC loading effects from the parasitic interconnects. To verify the robustness of programming against write disturb, we show the SPICE simulation results of the polarization states of a selected, half-selected, and unselected cell in the 32×32 array under the standard $V_{DD}/2$ write scheme (**Figure 6c**, right). The polarization switching transients confirm that target cell can be successfully programmed while the half-selected cells maintain a negligible polarization of 1.4%. This circuit modeling platform effectively evaluates the performance of both the material and the circuit architecture.

## 3. Conclusion

While the thriving of complex atomic-scale computational methods has greatly enriched fundamental understanding of ferroelectric switching mechanisms[31,42–44], the DFNG model presented in this work is compact yet able to serve as a complementary framework that links physical parameters related to microscale structure to experimental observables such as realistic switching transients, domain wall motion, and domain size. It enables extraction of material parameters under arbitrary time-dependent waveform and fast evaluation of technology metrics. We envision the model to be a powerful tool for performance-guided materials design and circuit-level optimization, thereby accelerating the development of next-generation ferroelectric computing technologies.

## 4. Methods

*Thin Film Growth*: The $Hf_{0.5}Zr_{0.5}O_2$ (W 20 nm/HZO 10 nm/W 50 nm) film was grown using plasma-enhanced atomic layer deposition on an undoped (001)-oriented silicon substrate, with the bottom and top tungsten layers deposited by DC magnetron sputtering[45]. The film was crystallized at 600 °C for 30 seconds via rapid thermal annealing in a dynamic $N_2$ atmosphere at atmospheric pressure. The epitaxial $BaTiO_3$ ($SrRuO_3$ 20 nm/BTO 20 nm/$SrRuO_3$ 50 nm) was grown by pulsed laser deposition on a (110)-oriented $DyScO_3$ single crystal substrate. The $Al_{0.92}B_{0.08}N$ (TiN 50 nm/AlBN 10nm/Ti 100 nm) film was grown on low-doped (100) Si via RF reactive magnetron co-sputtering[46] .

*Ferroelectric Capacitors Fabrication*: The HZO and BTO ferroelectric capacitors were fabricated using a combination of E-beam lithography and optical photolithography. A negative ma-N 2405 E-beam resist was spun on the trilayer film stack and exposed using a JEOL JBX-6300FS electron beam lithography system to serve as an etch mask to define the capacitor size. Then the sample was etched down to the bottom electrode with ion mill (Nanoquest II) to create self-aligned top electrode / ferroelectric islands. The sample was backfilled with 140 nm $SiO_2$ using magnetron sputtering (PVD 75 Proline, Lesker) for low κ isolation between the top and bottom electrodes. Next, a series of photolithography steps were performed to pattern bottom electrodes and lift-off ground planes. Finally, gold (300 nm)/titanium (10 nm) top contact pads were deposited with an E-beam evaporator. The AlBN capacitors were fabricated by first depositing Pt top contacts with a shadow mask and then etching the TiN layer with 30% $H_2O_2$.

*Ultra-fast Ferroelectric Switching Measurements*: Ultra-fast measurements were performed in an RF probe station. Ground-Signal-Ground probes (Model 40A, GGB) and low loss RF cables (SF526S/11PC35, HUBER+SUHNER) were used to transmit waveforms to and receive current response from the device. The voltage pulse train was generated by a pulse generator with a rise time down to 100 ps, connected to the top electrode, while the signal was recorded with a 13 GHz oscilloscope with a 50 Ω input, connected to the bottom electrode. With this setup, the current in the series RC circuit can be determined using the oscilloscope 50 Ω internal loading. For real-time voltage monitoring, the high-impedance probes were landed at the top and bottom electrodes to capture the differential voltage across the capacitor. All measurements were conducted in real-time with averaging over three single-shot measurements. To probe the ferroelectric switching transients, a reset pulse with negative polarity was first applied, followed by a positive-up (PU) pulse train. Each pulse width was 1 μs. The ferroelectric current was then extracted by subtracting the current during the U pulse (dielectric current and resistive current) from the current during the P pulse (ferroelectric current, dielectric current and resistive current).

*Numerical Model and Fitting*: Equation 1, 2, 3, 5 and S3 in Supplementary were used to generate $f$, $df/dt$ as a function of time $t$ and $\ln(-\ln(1-f))$ as a function of $\ln t$. Time and real-time voltage involved in the equations were experimental values. The slope of the dataset ($\ln(-\ln(1-f))$, $\ln t$) is the classical Avrami exponent. The sum of squared errors between model-generated outputs and experimental measurements under four applied voltage waveforms was minimized using a differential evolution algorithm to obtain the global optimum parameter set. After convergence, the Jacobian matrix of residuals with respect to the fitted parameters was evaluated and the associated covariance matrix was obtained via nonlinear least-squares refinement. The fit error was determined as the square root of the diagonal elements of this covariance matrix. We added a weight of 2 to the data points during 0-500 ps for HZO and BTO, as the experimental data points are taken every 7 ps and the experimental voltage rise time is ~200 ps, to reduce the intrinsic bias of the least squares fit at low time and $f$ values. This weight was chosen to not overbias this region as well. For AlBN, a weight of 10000 is added to the $df/dt$ data, as the original values are 10000 times smaller than the other datasets. This weight is to ensure all datasets contribute comparably to the overall fit. An analysis of the weighing method is provided in **Supplementary Note 7** and **Figure S8**.

## Figures and Captions

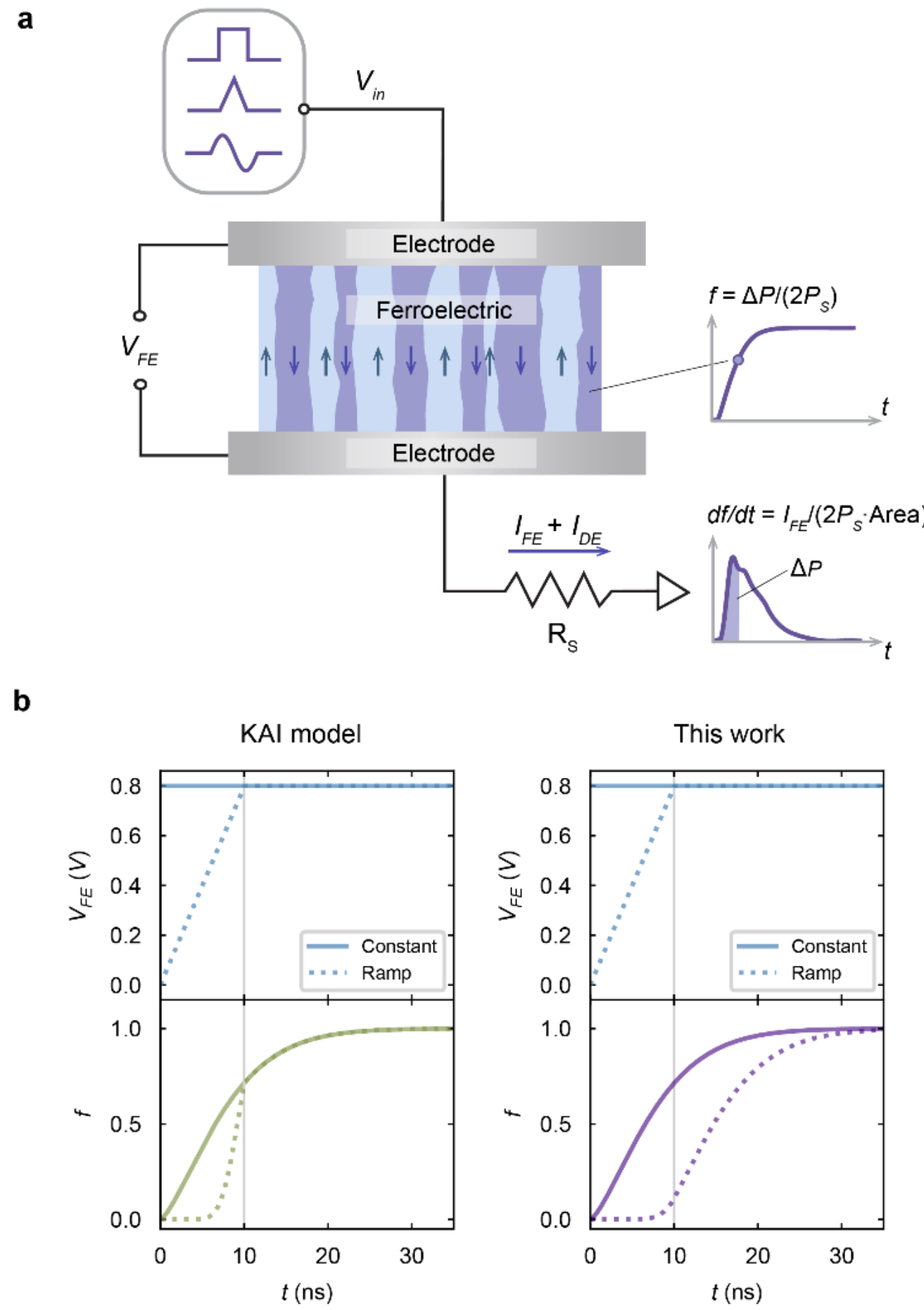

**Figure 1. Ferroelectric switching transients under dynamic electric fields. a,** A schematic of a ferroelectric capacitor driven by various time-dependent voltage waveforms ($V_{in}$). Assuming the ferroelectric material is preset to an up-polarized state (blue) before $t = 0$, the waveform induces partial switching via the nucleation and growth of down-polarized domains (purple). The voltage across the ferroelectric capacitor ($V_{FE}$) and the current transient, including ferroelectric switching current ($I_{FE}$) and linear dielectric current ($I_{DE}$) are measured. By integrating $I_{FE}$ and normalizing it to spontaneous polarization ($P_S$), the instantaneous switched (transformed) fraction ($f$) can be determined. **b,** Illustration of the error from a naive voltage-dependent KAI model contrasted with the DFNG model presented here. Given a constant applied voltage (solid lines) and voltage ramp to the same set point (dashed lines). The KAI model produces a polarization transient insensitive to the ramp history, in contrast to the DFNG model.

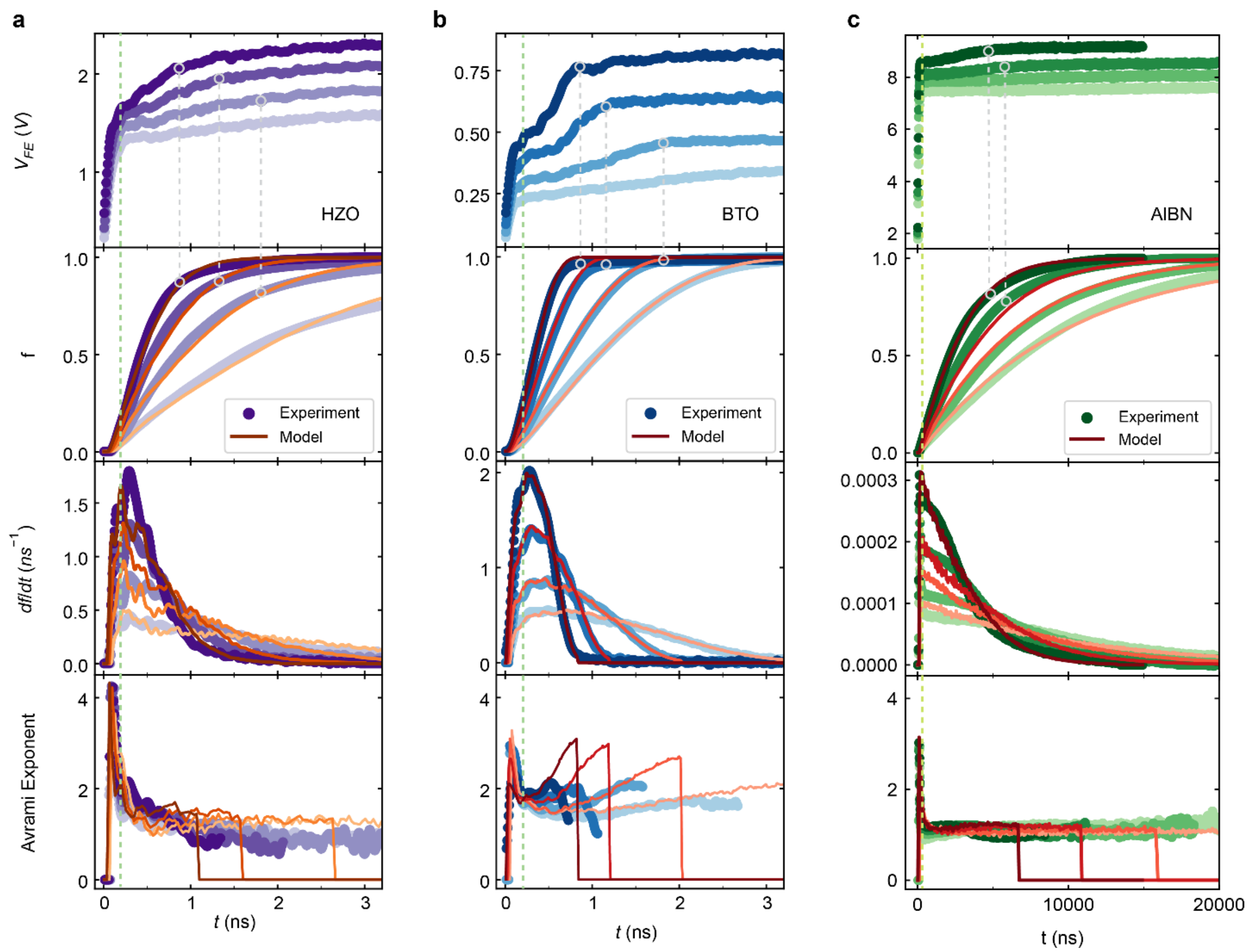

**Figure 2: Self-consistent fitting of the DFNG model across distinct material systems and various dynamic waveforms.** Experimental $V_{FE}$, $f$, normalized $I_{FE}$ ($df/dt$), and Avrami exponent over time of a polycrystalline HZO (**a**), single-crytalline BTO (**b**) and wurtzite AlBN (**c**) capacitors along with the DFNG model fit.The green dashed vertical line marks the end of the fast voltage ramp of the supply (200 ps for HZO and BTO, 100 ns for AlBN). The region between the green and grey dashed lines indicates waveform distortion due to a circuit effect. The grey circles mark the corresponding transformed fraction when the voltage saturates to the setpoint. Notably, most switching occurs under a time-varying voltage beyond the validity of traditional nucleation and growth models. The DFNG model accurately reproduces these transients and captures the evolving Avrami exponent, enabling physical interpretation beyond fixed growth dimensionality.

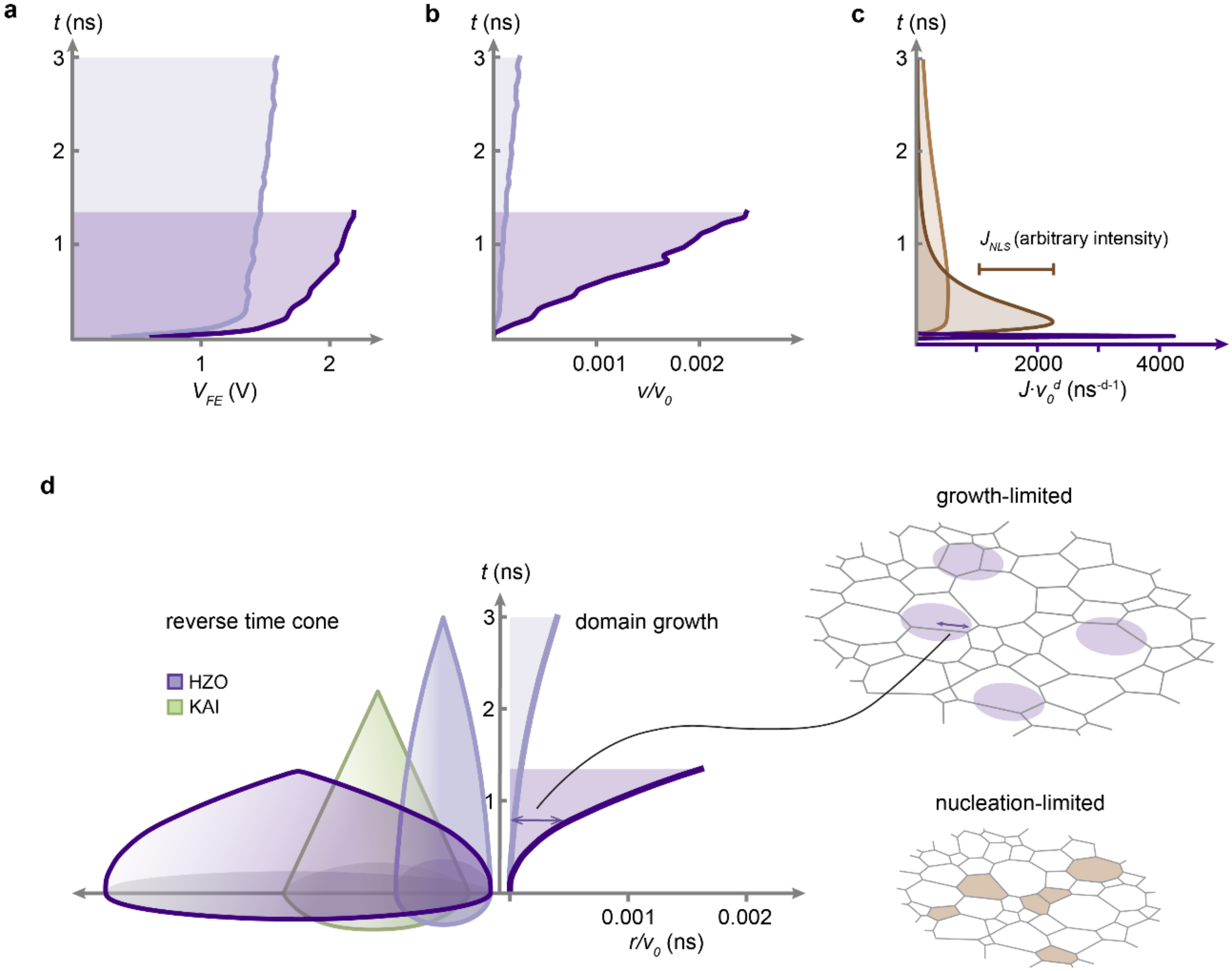

**Figure 3. Time-dependent voltage driven domain nucleation and growth. a,** Representative experimental time-dependent voltage waveforms across the HZO ferroelectric capacitor. The darker curves deviate more from a square pulse than the lighter curves due to the more pronounced circuit effect. **b,** Extracted domain wall velocity (normalized). The domain wall velocity of HZO roughly increases linearly in time. **c,** Extracted nucleation rate (normalized) of HZO in comparison with a qualitative nucleation profile derived from the NLS (nucleation-limited switching) model. **d,** Extracted reverse time cones and transformed domain radii. In real materials, the time-varying domain wall velocity leads to a curved lateral surface of the reverse time cone. In the idealized KAI model (green), the cone has a straight profile due to its limiting assumptions. Our model reveals that the transformed domain can grow to a radius of a few nanometers, which is comparable to the grain size, but is not restricted inside a grain (top right schematic). In contrast, the NLS model assumes that the domain wall is stagnated by the grain boundary (bottom right schematic). The colored areas are the transformed regions.

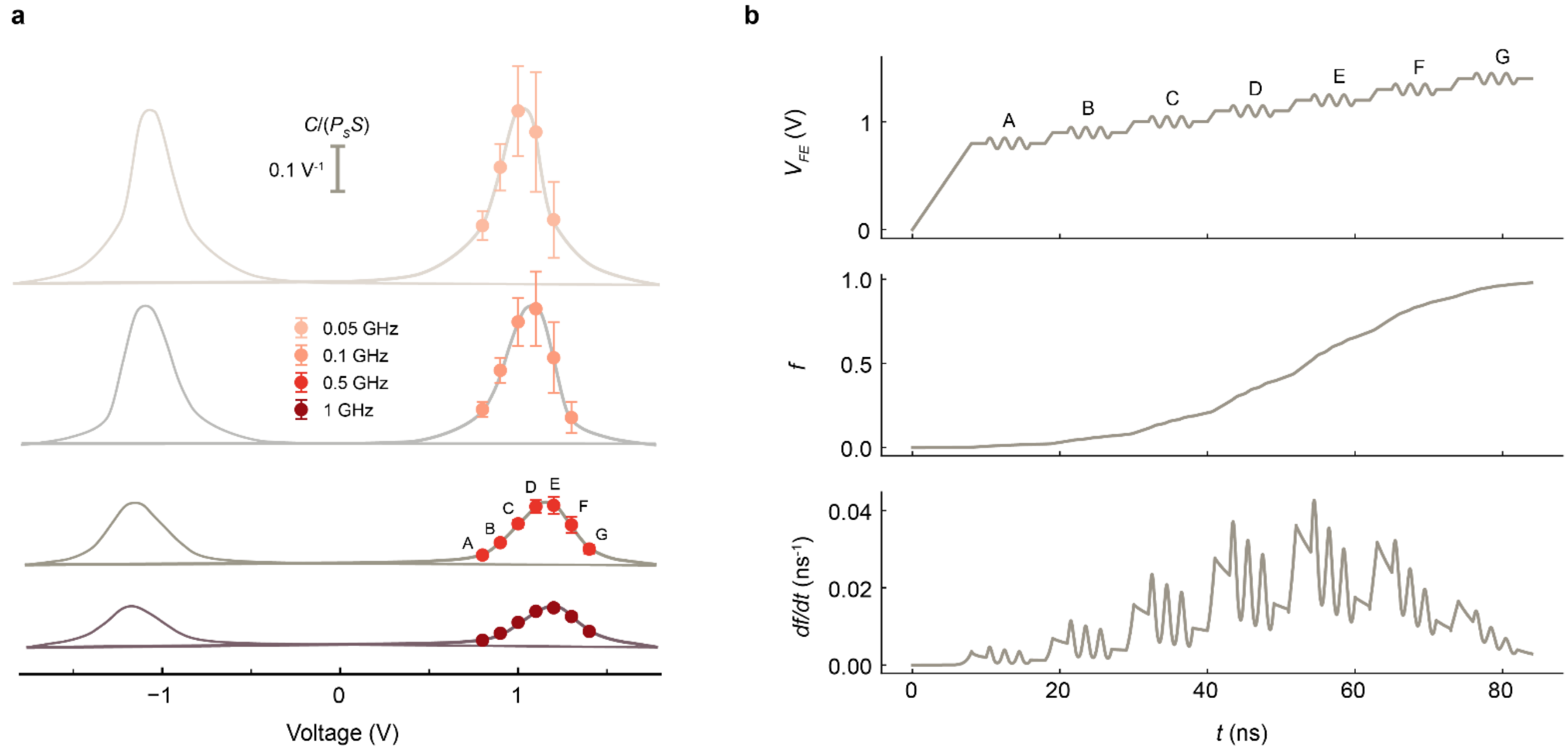

**Figure 4. Mixed-signal simulation: small-signal capacitance. a,** Simulated frequency-dependent capacitance-voltage $C(V)$ hysteresis. Red dots with error bars are output from the model. Grey curves are guide to eyes. **b,** Assumed waveform ($V_{FE}$) used to acquire peak capacitance, transformed fraction ($f$), and normalized current ($df/dt$), corresponding to $C(V)$ data points at 0.5 GHz in (**a**). The first three oscillations in the voltage and current are used to generate mean capacitance values and standard deviations. The model captures $C(V)$ hysteresis as well as empirically known behavior such as the decrease in the DC voltage at peak capacitance with decreasing frequency and decreasing peak capacitance with increasing AC frequency. The model also illustrates that polarization drift may lead to significant variance in $C(V)$ data. The model can handle arbitrary mixed voltages spanning orders of magnitude in time and frequency scales.

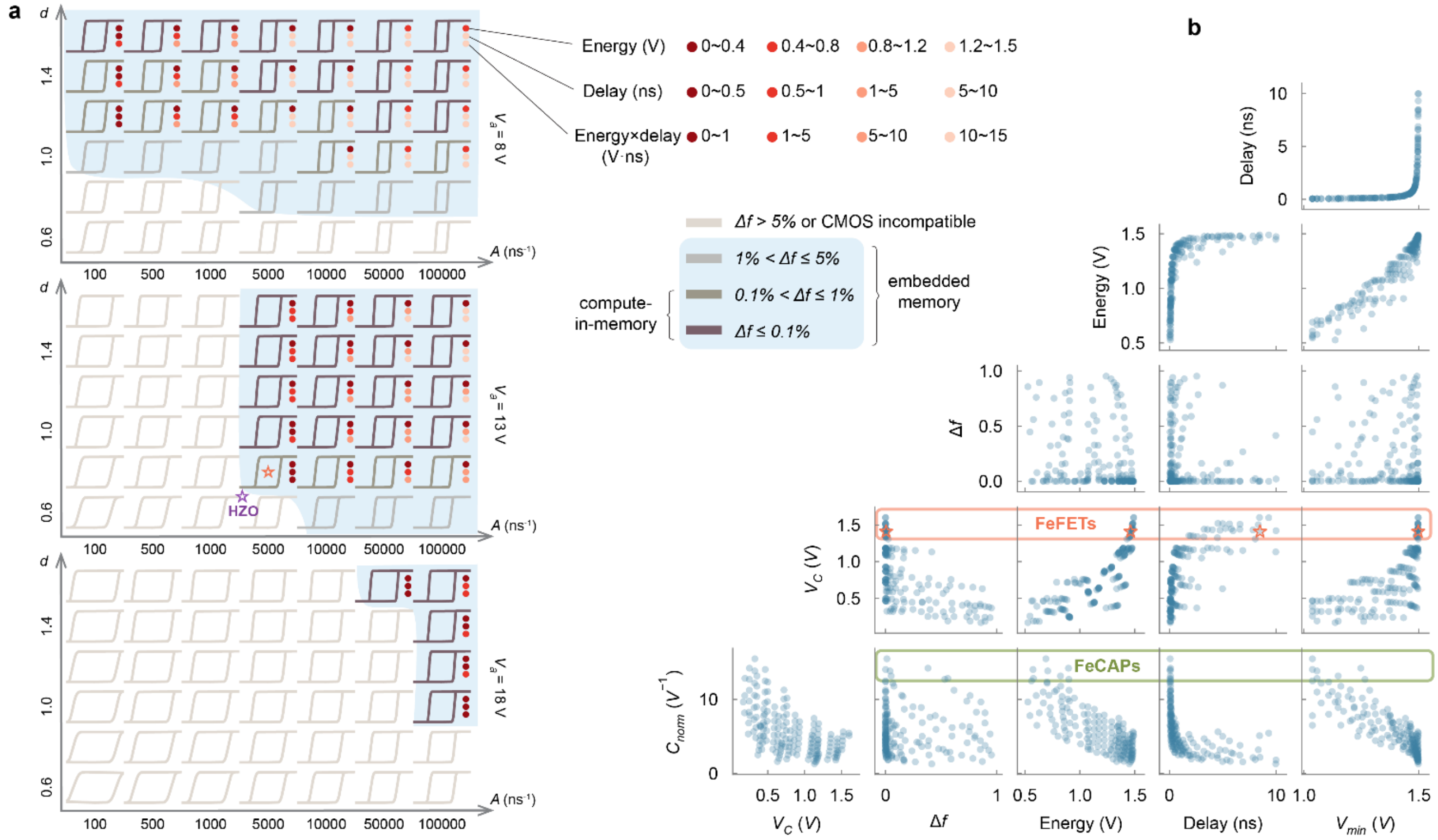

**Figure 5. Mapping material design space to application metrics. a,** Hysteresis loops generated in a segment of model parameter space using a 1 MHz triangular waveform. The same material parameter sets are used to extract application metrics. We grade each loop according to nominal CiM and embedded memory application standards, illustrated in the legend. The values for energy and capacitance are normalized by polarization and capacitor area. **b,** The correlation between application metrics for CiM devices. The orange box selects parameter sets that render a relatively large $V_C$ and hence sufficient read margin for FeFETs. The green box selects parameter sets that allow a relatively large $C_{norm}$, which is the memory window in FeCAP devices. Orange stars in (**b**) label a representative parameter set suitable for FeFET and correspondingly in (**a**) indicates a possible modification direction of the current HZO sample (purple star) for such application. However, modifying it for FeCAPs require demanding conditions and hence may not be ideal. Notations used: $\Delta f$: normalized polarization loss due to $V_{DD}/2$ write disturb; $C_{norm}$: normalized peak differential capacitance, calculated by $\left(\frac{df}{dV}\right)_{f=0.5}$; $V_C$: coercive voltage; $V_{min}$: minimum pulse amplitude to reach 99% switched fraction within 10 ns.

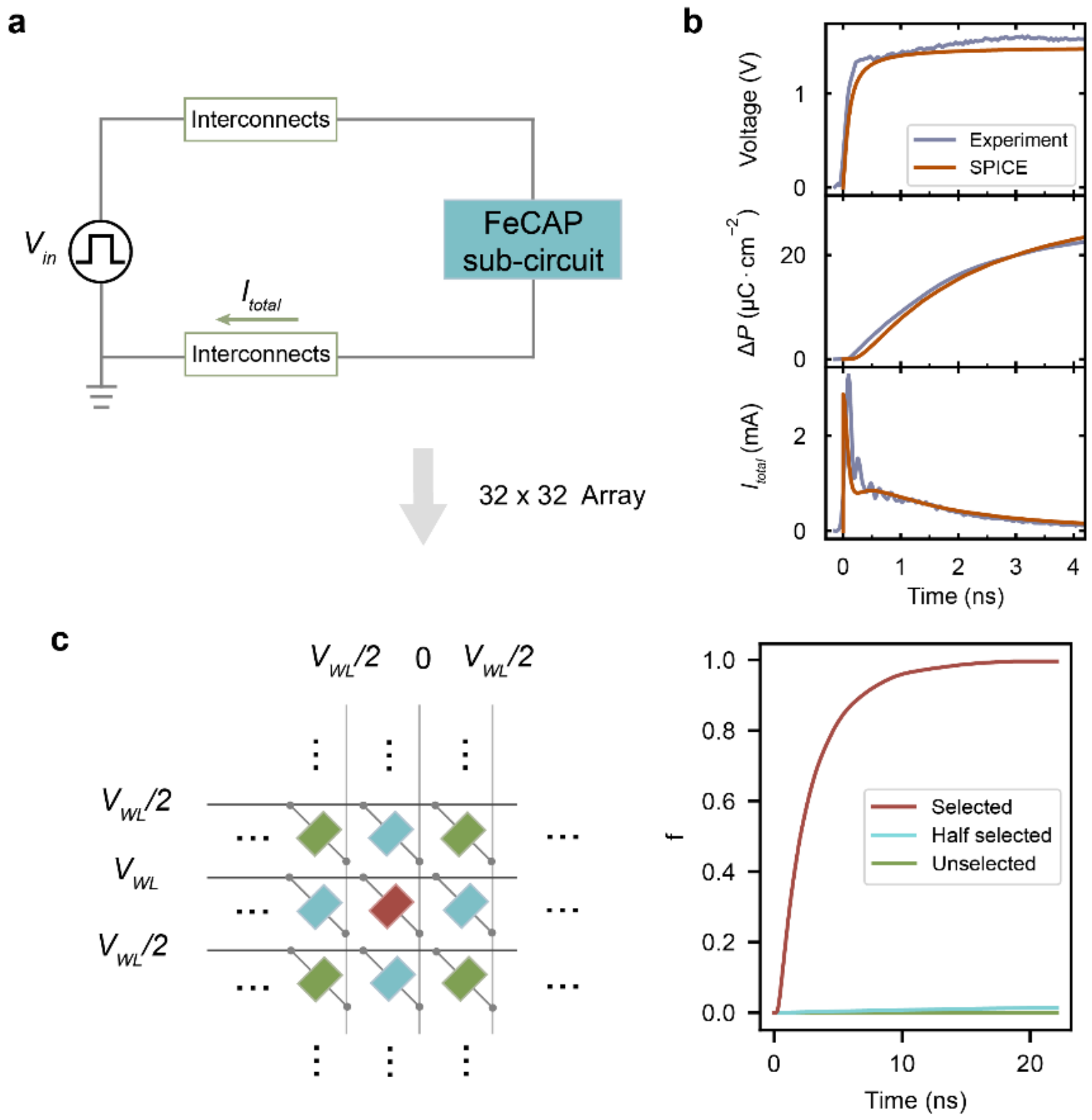

**Figure 6**: **Scalable circuit modeling platform with physical material parameters. a**, A schematic of the equivalent circuit of a single ferroelectric device implemented in sub-circuits in SPICE. The FeCAP SPICE model accurately implements the DFNG model to simulate the dynamics ferroelectric polarization, as discussed in detail in Supplementary Note 6. **b**, Polarization and total current transients measured experimentally (purple) and simulated by the FeCAP SPICE model (red). **c**, Schematic of a 32×32 FeCAP crossbar array under the $V_{DD}/2$ write scheme (left) and the polarization transients of the selected, half selected and unselected cells in this array (right) produced by SPICE simulations, taking account realistic parasitic effects in the crossbar. The SPICE simulation verified that the selected cell can be successfully programmed while the half-selected cell and unselected cells maintain negligible polarization (<1.4%).

**Acknowledgements**

Y.L., T. C., and J. T. H. acknowledge partial support from the Intel FEINMAN 2.0 program. Y.L. and J. T. H. additionally acknowledge partial support from the ONR grant N000142612047. This work was performed in part at the University of Michigan Lurie Nanofabrication Facility. HZO films were prepared under support from the Laboratory Directed Research and Development Program at Sandia National Laboratories. Sandia is a multimission laboratory managed and operated by National Technology and Engineering Solutions of Sandia LLC, a wholly owned subsidiary of Honeywell International Inc. for the U.S. Department of Energy's National Nuclear Security Administration under Contract No. DE-NA0003525.

**Data Availability Statement**

All data that supports the findings of this study are available from the corresponding author on reasonable request.

## Supporting Information

### 1. Deficiency of KAI model in capturing switching behavior under time-dependent voltage

KAI model describes the transformed fraction *f* of polarization at time *t* by:

$$f(t) = 1 - \exp\left[-\left(\frac{t}{t_0}\right)^n\right] \quad (S1)$$

where $t_0$ is the characteristic time related to voltage-independent nucleation rates and domain wall velocities, and *n* is the Avrami exponent related to domain growth dimension. The conventional form of the KAI model only applies to the constant electric field case. To involve voltage dependence, Merz law can be introduced to $t_0$, i.e. $t_0(t) = t_\infty \exp\left[\frac{V_a}{V(t)}\right]$, so that

$$f(t) = 1 - \exp\left[-\left(\frac{t}{t_\infty \exp\left[\frac{V_a}{V(t)}\right]}\right)^n\right] \quad (S2)$$

$t_\infty$ is the characteristic time at infinite voltage, $V_a$ is the activation voltage of domain wall depinning, and $V(t)$ is the instantaneous voltage across the ferroelectric capacitor. With the voltage-dependent KAI model, the polarization response under the ramp (Figure 1b, green dashed line) coalesces with that under constant bias (Figure 1b, green solid line) once the set point is reached. This behavior is unphysical, as the applied voltage during the ramp remains below the set point prior to 10 ns, and the polarization would therefore be expected to require a longer time to reach the same value. Therefore, a model that captures the voltage path dependence is needed to correctly reflect the polarization switching physics.

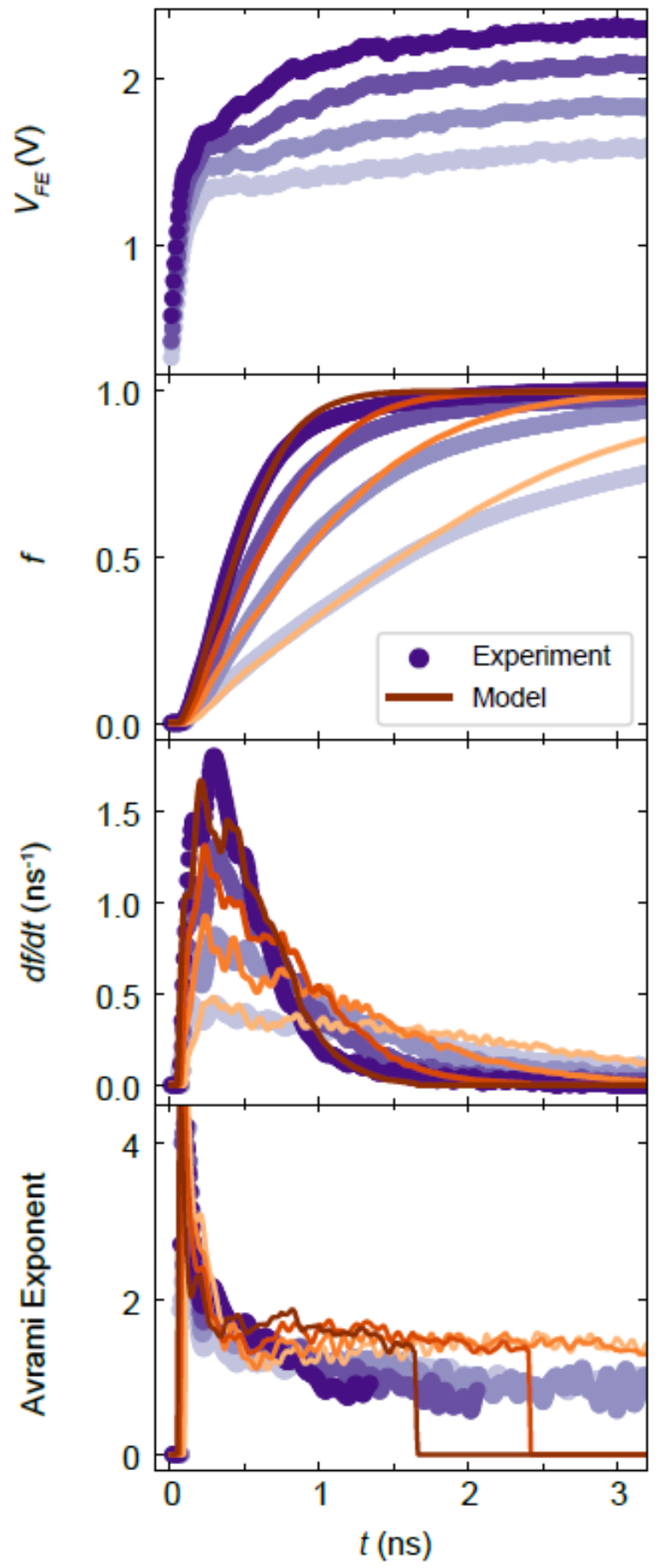

**Figure S1: Fitting HZO transients with both heterogeneous and homogeneous nucleation.** The experimental transients of a 3 µm diameter HZO capacitor is fit to the model using Equation (S5) with 5 parameters, including heterogeneous and homogeneous nucleation terms. The model yields parameters $d = 0.808 \pm 0.004$, $V_a = 11.68 \pm 0.12\ V$, ${\sigma'}^2 = 42.87 \pm \infty\ V$, $A = 1918 \pm 138\ ns^{-1}$, and $B = 1587 \pm \infty\ ns^{-\frac{1}{d}-1}$. Despite that the modeled data (orange) matches relatively well to the experimental data (purple), the model cannot produce a converged solution to parameter ${\sigma'}^2$ and $B$. Moreover, the unreasonably high domain wall energy indicates that homogeneous nucleation is unlikely in the HZO system. Therefore, the polarization switching in HZO is solely dominant by heterogeneous nucleation.

## 2. The extraction of material parameters for BTO

Following the fit procedure in Methods, the experimentally measured switching transients of a 3 µm BTO capacitor is fit to the model, as shown in **Figure 2b**. The data best fits the model with a nucleation rate including both heterogeneous and homogenerous nucleation, i.e. summing Equation 3 and 5. Therefore, the total $\langle N(t) \rangle$ is given by:

$$\langle N(t) \rangle = A^d \left[ \int_0^t \exp\left[ -\frac{V_a}{V(t')} \right] dt' \right]^d + B^d \int_0^t \exp\left[ -\frac{V_a + \sigma'^2}{V(\tau)} \right] \left[ \int_\tau^t \exp\left[ -\frac{V_a}{V(t')} \right] dt' \right]^d d\tau \qquad (S3)$$

The extracted parameters are $d = 1.144 \pm 0.004$, $V_a = 0.536 \pm 0.003\ V$, $\sigma'^2 = 1.47 \pm 0.04\ V$, $A = 6.13 \pm 0.07\ \mathrm{ns}^{-1}$, and $B = 713 \pm 44\ \mathrm{ns}^{-\frac{1}{d}-1}$. Specifically, the Avrami exponent of BTO shows a distinct behavior compared to HZO, which comes from the contribution of homogeneous nucleation. The universal applicability of the model under dynamic field is manifested by good fits to the switching transients of both polycrystalline materials and single crystalline materials under various distorted waveforms. The reduction of the growth dimension *d* from the physical device dimensions (2D) can be attributed to impediment of the domain wall motion in the system such as the finite size effects, grain boundaries and defects, which produces a truncated time cone[1]. The truncated time cone also accounts for the deviation between the model and the experimental data when approaching the end of the switching. Besides a higher growth dimension compared to HZO, BTO also exhibits a much lower Merz barrier ($V_a$) and heterogeneous nucleation density ($A$). The domain wall energy $\sigma$ of BTO is derived to be ~4.4 mJ/m$^2$, according to $\sigma'^2 = \frac{\pi t_{FE}^2 \sigma^2}{2 P_S k_B T}$, and using $P_S = 20\ \mu\mathrm{C/cm}^2$, $t_{FE} = 20$ nm, $T = 300$ K. The domain wall energy derived by our model is comparable to *ab initio* calculated values[2]. Homogeneous nucleation density is estimated as ~$10^{-4}$ nm$^{-d}$ using a dipole attachment frequency $\omega_0$ of 1 THz[3], and $v_0$ of 5000 m/s[4]. It reveals that nucleation in BTO is sparse and that the domains are large (100s-1000s nm).

### 3. Material parameters extraction in AlBN

Due to the limitation of the experimental setup, which can only handle votlages up to 12 V, 10 nm thick AlBN thin film is selected in this study, which only requires a switching voltage of 7-9 V. The experimental data best fit a 3-parameter DFNG model with heterogeneous nucleation. The extracted parameters are $d = 0.992 \pm 0.004$, $V_a = 66.0 \pm 0.6$ V, $A = 0.71 \pm 0.06\ \mathrm{ns}^{-1}$. The experimental data and model fit results are shown in **Figure 2c**.

We note that the leakage current is non-negligible in this ultrathin sample as well as those widely reported in the literature[5,6]. The leakage current can be nonlinearly convoluted with voltage and polarization[7], as it cannot be fully accounted for by subtracting the U pulse current in a standard PUND measurement (**Figure S2**). To the best of our knowledge so far, there is no established approach in the community to handle the correlated leakage easily and accurately. Therefore, we use an approximation (**Figure S3**) where the leakage current is forced to saturate to a constant value with a linear ramp from zero, so that the ferroelectric switching current ($I_{FE}$) can decay to zero at the end of the switching process. Specifically, we first calculate a current ($I_P - I_U$) by subtracting the U pulse current from the P pulse current. Then the peak position of the second derivative of this current ($I_P - I_U$) determines the end of the linear ramp of the leakage current (Figure S3a), because the maximum curvature in the current transient may indicate a transition from the ferroelectric displacive origin to some domain relaxation/ domain wall conduction origin. The static value of the leakage current is taken at a time equal to three times the peak position. With this method, the saturation polarization is consistently saturate to comparable values (155-165 $\mu\mathrm{C} \cdot \mathrm{cm}^{-2}$) across different supply voltages. The leakage level is much smaller than the ferroelectric switching current (≤ 10%). As a result, though different subtraction methods may affect the exact model fit parameters, they are unlikely to change the order of magnitude of the physical variables and the projected mechanism.

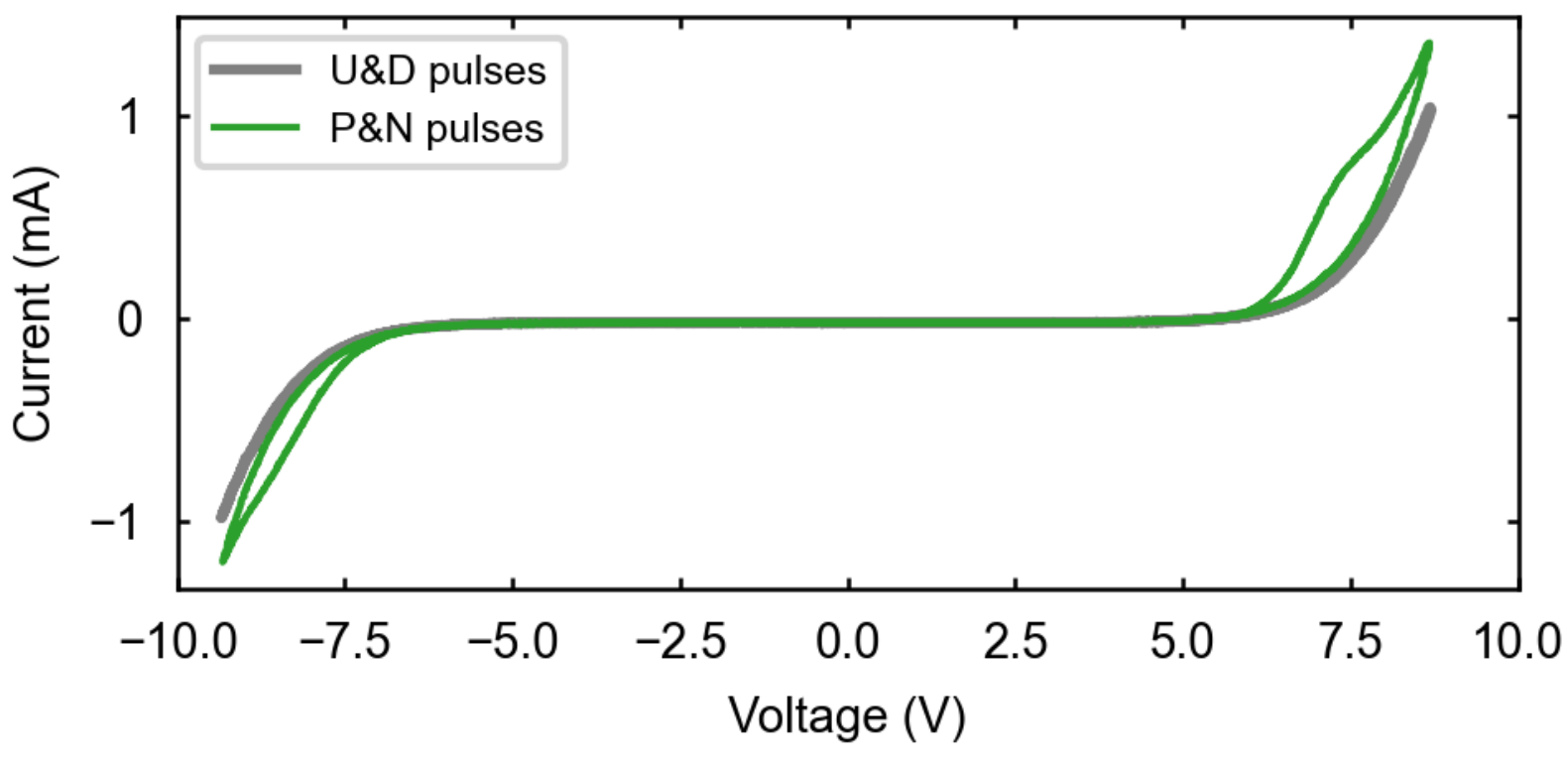

**Figure S2: Leakge current in ultrathin AlBN.** The leakage current is nonlinearly convoluted with voltage and polarization.

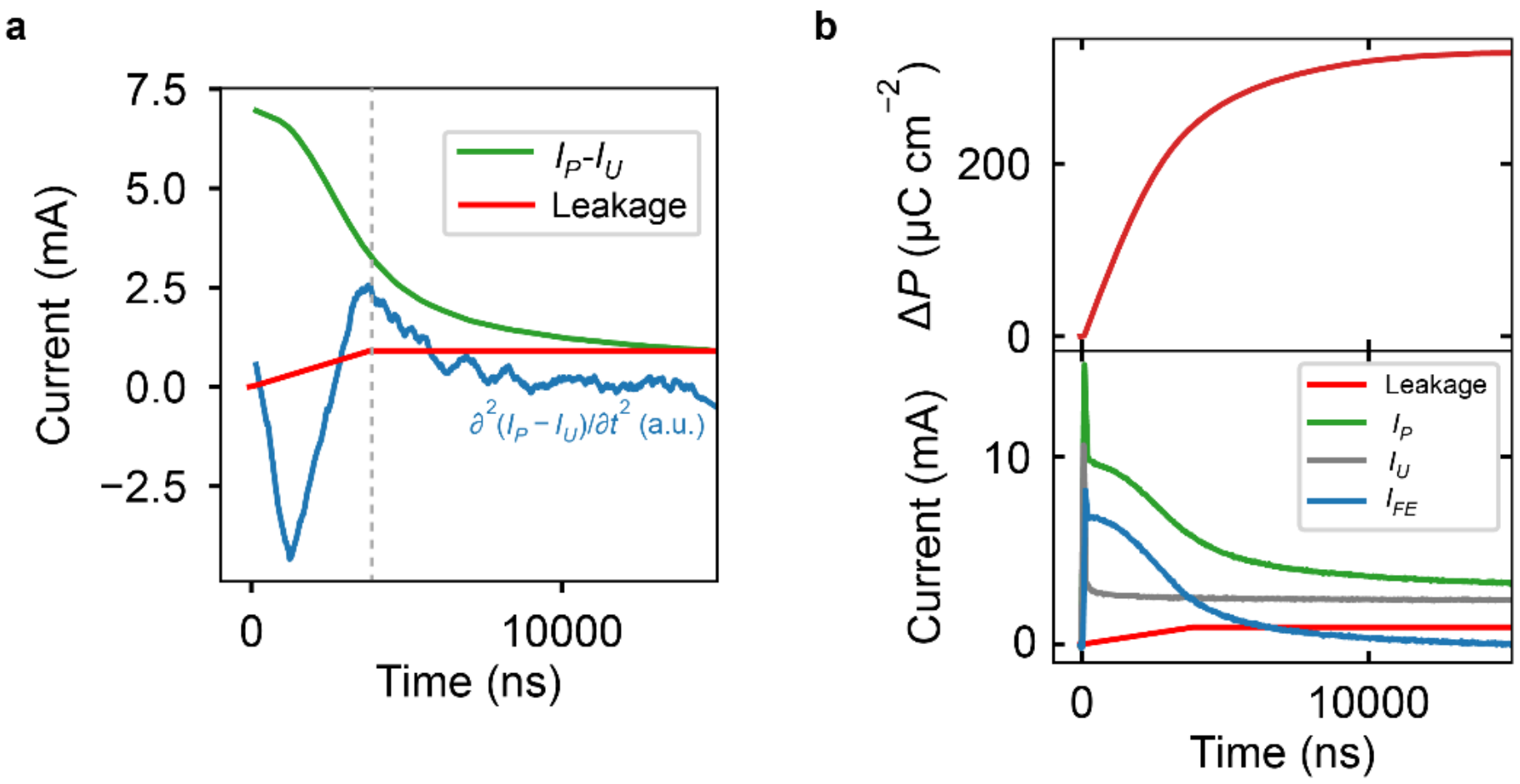

**Figure S3: Leakage subtraction. a,** The leakage current is assumed to be a linear ramp with a duration determined by the peak position of the second derivative of $(I_P - I_U)$, and to a static value determined at a time equal to three times the peak position. **b,** With the leakage subtracted, the ferroelectric current can decay to zero and polarization can saturate.

## 4. Parameter space exploration and application metrics extraction

### 4.1 Simulation of hysteresis loops

Leakage, linear dielectric constant, and depolarization in the ferroelectric capacitor are ignored and a preset -$P_S$ state is assumed. The method to generate the polarization switching during a voltage ramp is shown in **Figure S4a** and **S4b**, which is essentially half of a hysteresis loop. The voltage ramp is described by $V_{FE}(t) = V_0 \cdot \frac{1}{T/4} \cdot t$. $T$ is the period of the hysteresis loop, and the frequency $F$ of the loop is given by $1/T$. $V_0$ is the set voltage of the ramp. The frequency of the loops in **Figure 5a** is 1 MHz, and the set voltage is 3 V. The full hysteresis loop is shown in **Figure S4c**. The polarization only switches during 0-250 ns and 500-750 ns per the assumption, and the switching transients in these two periods are asymmetric. Interestingly, this method can be used to generate frequency dependent hysteresis loop. The frequency dependent coercive voltage ($V_C$) matches the empirical power law $V_C \propto F^{\beta}$ [8](**Figure S5**).

For parameter space exploration, we iterate every combination of parameters among $d$ = 0.6, 0.8, 1.0, 1.2, 1.4, 1.6, 1.8; $V_a$ = 4, 8, 13, 18 V; $A$ = 10, $10^2$, $5 \times 10^2$, $10^3$, $5 \times 10^3$, $10^4$, $5 \times 10^4$, $10^5$ $ns^{-1}$.

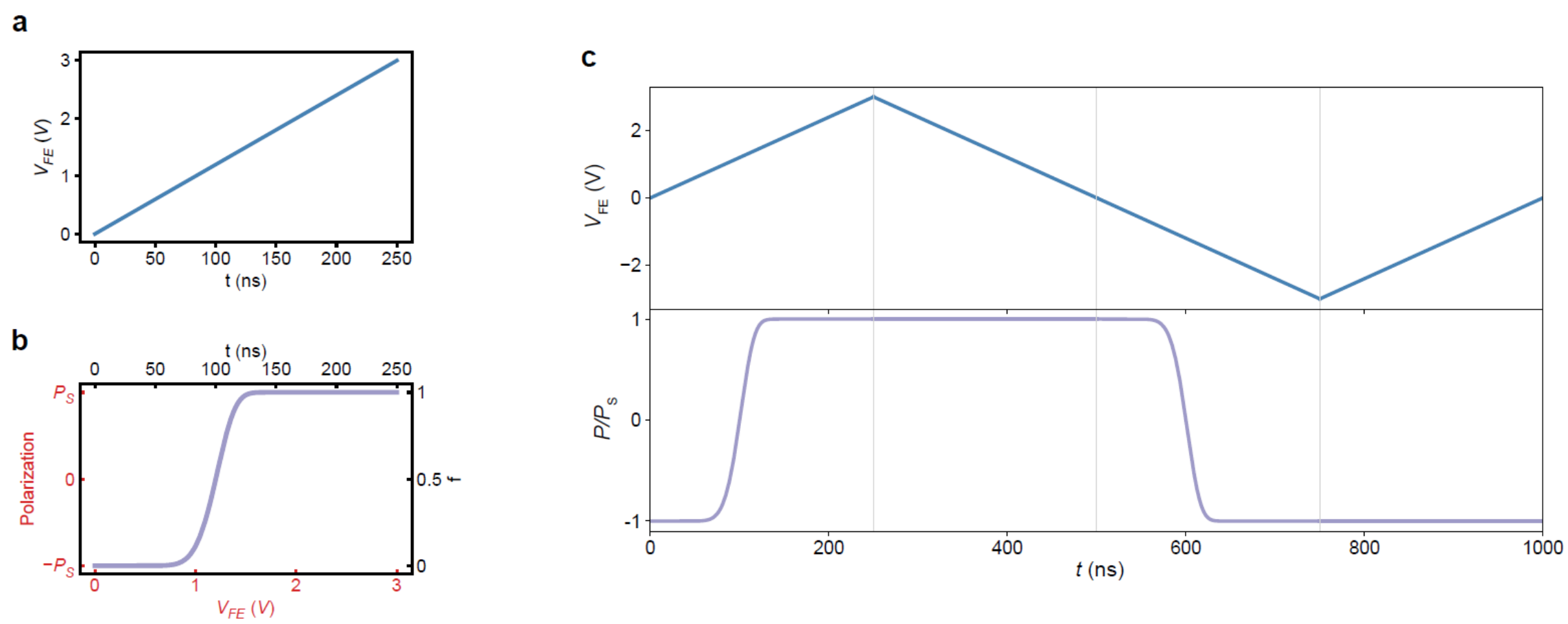

**Figure S4: Hysteresis loop generation. a,** The voltage ramp of a 1 MHz hysteresis loop. **b,** The transformed fraction $f$ as a function of time can be mapped to the polarization as a function of voltage. The ferroelectric polarization is preset to $-P_S$, and switches to $P_S$ during the voltage ramp, corresponding to a transformed fraction from 0 to 1. **c,** The full waveform of a hysteresis loop measurement and the corresponding polarization values.

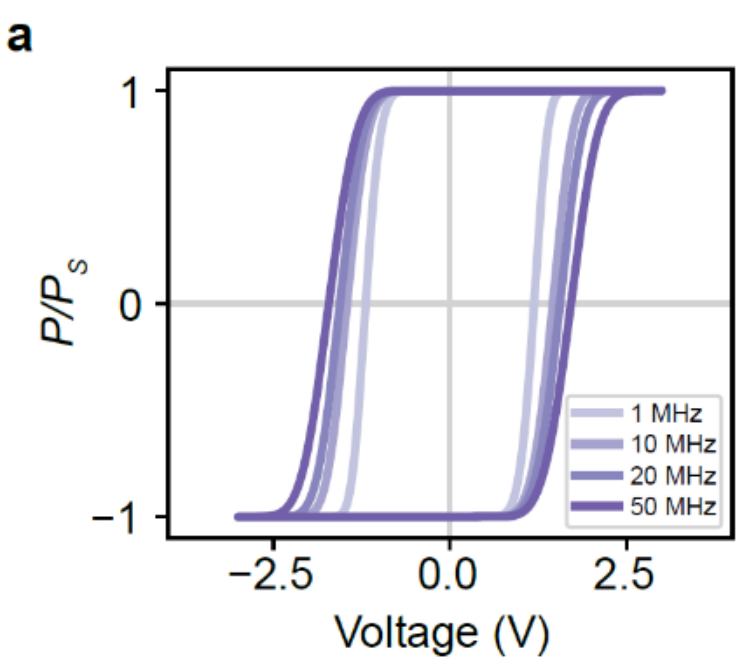

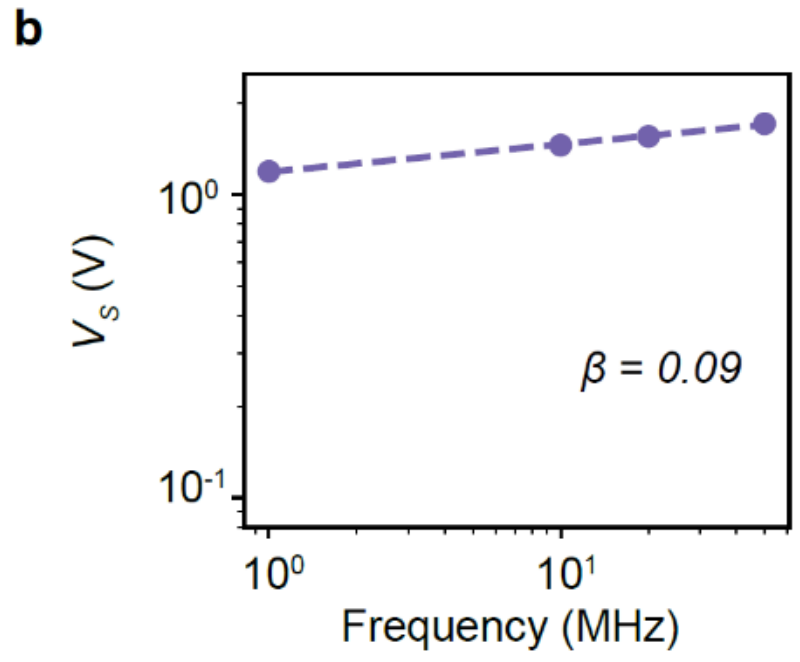

**Figure S5: Frequency dependent hysteresis loops. a,** Simulated hysteresis loops under 1, 10, 20, 50 MHz triangular waveform. **b,** Frequency dependent coercive voltage. Using the parameter sets of the 3 µm HZO capacitor, the model captures the empirical power law $V_C \propto F^{\beta}$. The dots are the simulated coercive voltages, and the dashed line is a linear fit. The linear fit gives a $\beta$ value of 0.09.

**4.2 Extraction of application metrics and grading of the loops**

Instead of triangular waves and linear ramp of voltages, pulse waveforms are more commonly used in circuits operations for higher latency. All simulated pulse waveforms used in this work are defined as $V_{FE}(t) = \frac{2}{\pi} V_0 \arctan 50t$. $V_0$ is the pulse amplitude. This equation gives a rise time of ~80 ps to reach ~0.85$V_0$. A simulation example is shown in **Figure S6**.

To grade the loops according to the procedure mentioned in the main text, we first simulate the switching transient under a pulse operation of 1.5 V for every parameter set involved. Then we select the parameter sets that allow more than 99% switching fraction at 10 ns. The voltage when $f = 0.99$ is minimum pulse amplitude required ($V_{min}$). We run the simulation again under a pulse with an amplitude of $V_{min}$ to extract the delay, which is the time needed to complete 99% switching transformation, as well as under a pulse with an amplitude of $V_{min}/2$ and a width of the delay to extract $\Delta f$ (normalized disturbed polarization in $V_{DD}/2$ write scheme).

For FeCap and FeFET devices, the key metrics are memory windows, namely the capacitance ($C_{norm}$) and coercive voltage ($V_C$), respectively. These two quantities are highly susceptible to the frequency of the electrical waveform, or equivalently the ramp rate of the voltage under a variety of application contexts. As a proof-of-concept, we extract these two values using a fast voltage ramp of 0.1 V/ns, which is approximate to a 320 MHz, 0.05 V AC small signal oscillation.

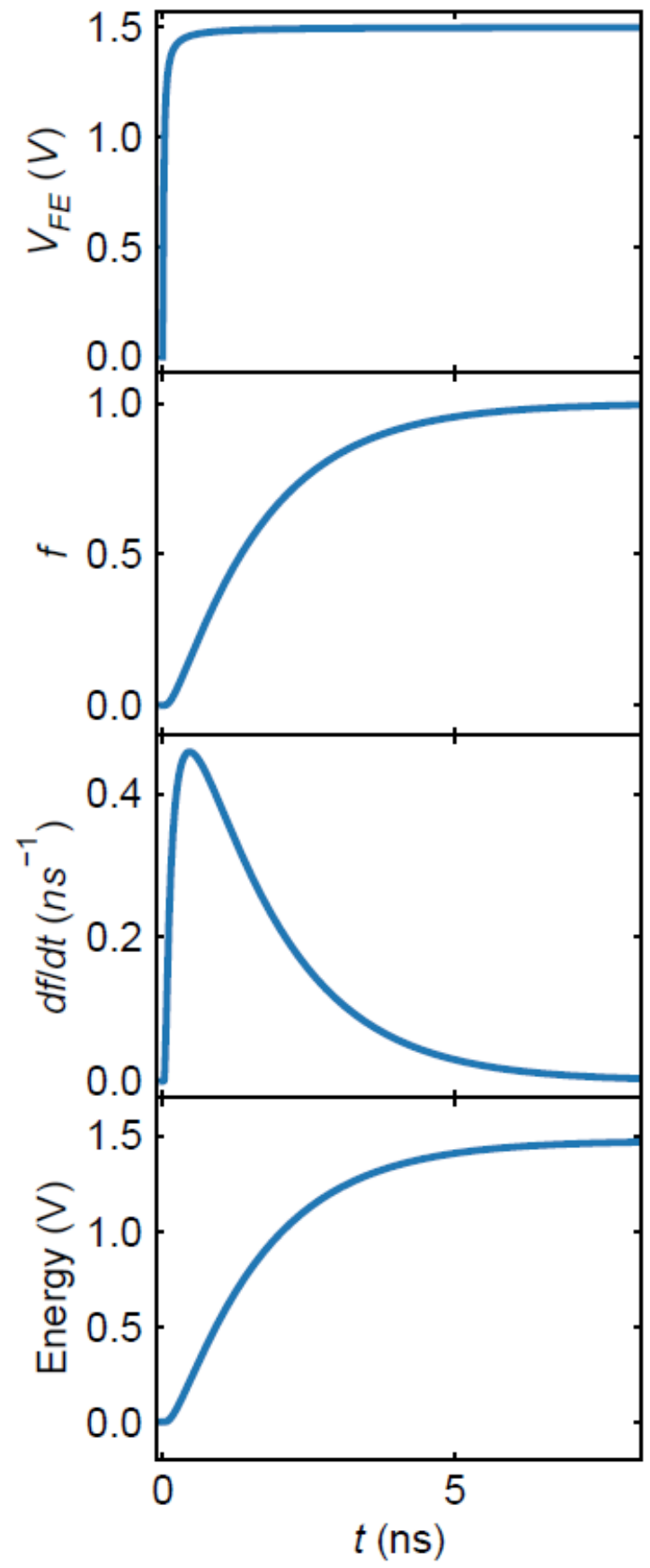

**Figure S6: Simulated switching transient under pulse waveform.** In this simulation, $V_0$ is set to be 1.5 V. After calculating the transformed fraction *f* and then the normalized ferroelectric current *df/dt*, energy is given by integrating the product of *df/dt* and $V_{FE}$ over time.

## 5. Parameter sets filtering for FeFETs and FeCaps

We use a nominal criterion of $V_C \geq 1.4$ V to select favorable FeFETs gate materials. It is inferred from **Figure 5b** that only the delay has a broad distribution and needs consideration under this selection. Table S1 is a table of the satisfying parameter sets, in the order of increasing delay. The parameter set that gives the minimum delay is highlighted in orange. The one that is closest to our measured HZO sample is highlighted in blue. It indicates a possible solution that if the depinning field is relatively unchanged, by slightly increasing the growth dimension (for example through geometric design or removing grain boundaries) and nucleation density (for example by introducing point defects), the HZO can be more suitable for FeFETs. Moreover, the parameter set highlighted in green also provides a potentially viable solution that by diminishing the pinning defects (and therefore increasing $d$, and reducing $V_a$ and $A$), HZO can be adapted to fit the device need.

$C_{norm} \geq 12$ $V^{-1}$ is selected for FeCap devices. From **Figure 5b**, this criterion gives small delay but a wide distribution in other metrics. If $\Delta f$ is required to be less than 1%, only three sets of parameters satisfy the standard (Table S2). These parameter sets have high growth dimension approaching the physical dimension (2D) and small depinning field, which is more aligned with single crystalline materials. However, the nucleation density needs to maintain high. The demanding values and the rarity of suitable parameters indicate polycrystalline materials are probably not ideal for FeCAPs.

| $d$ | $V_a$ (V) | $A$ ($ns^{-1}$) | Delay (ns) | $V_C$ (V) |
|---|---|---|---|---|
| 1.8 | 13 | 5000 | 3.25 | 1.44 |
| 1.6 | 13 | 5000 | 3.56 | 1.44 |
| 1.4 | 13 | 5000 | 4.01 | 1.44 |
| 1.8 | 18 | 100000 | 4.54 | 1.52 |
| 1.2 | 13 | 5000 | 4.73 | 1.43 |
| 1.6 | 18 | 100000 | 4.98 | 1.52 |
| 1.8 | 8 | 100 | 5.24 | 1.45 |
| 1.4 | 18 | 100000 | 5.62 | 1.52 |
| 1.6 | 8 | 100 | 5.79 | 1.45 |
| 1.0 | 13 | 5000 | 5.95 | 1.42 |
| 1.4 | 8 | 100 | 6.58 | 1.44 |
| 1.2 | 18 | 100000 | 6.61 | 1.51 |
| 1.2 | 8 | 100 | 7.83 | 1.43 |
| 1.0 | 18 | 100000 | 8.33 | 1.51 |
| 1.8 | 18 | 50000 | 8.44 | 1.60 |
| 0.8 | 13 | 5000 | 8.48 | 1.41 |
| 1.6 | 18 | 50000 | 9.30 | 1.60 |
| 1.0 | 8 | 100 | 9.98 | 1.42 |

**Table S1: Favorable parameters for FeFETs.**

| $d$ | $V_a$ (V) | $A$ ($ns^{-1}$) | $C_{norm}$ ($V^{-1}$) |
|---|---|---|---|
| 1.6 | 4 | 10000 | 12.51 |
| 1.8 | 4 | 5000 | 12.37 |
| 1.8 | 4 | 10000 | 14.06 |

**Table S2: Favorable parameters for FeCAPs.**

## 6. SPICE modeling implementation

The SPICE simulation was configured using an equivalent circuit model to capture the transient switching dynamics of the ferroelectric capacitor with interconnect parasitic effects. The setup assumes a HZO ferroelectric capacitor of 3 μm in diameter with the material parameters extracted by the DFNG model ($d = 0.699$, $V_a = 12.75$ V, $A = 3236\ \text{ns}^{-1}$), a saturation polarization ($P_S$) of $15\ \mu\text{C/cm}^2$, and an intrinsic linear capacitance of 300 fF. The ferroelectric capacitor is connected to a 3 μm wide Cu interconnect with 3 μm line space. Thus the interconnects produce a parasitic resistance of 0.4 Ω and a line-to-line capacitance of 6.207 fF. A square pulse of 1.5 V with an arctangent-shaped rising profile was employed as the input signal ($V_{in}$), $V_{in} = 1.5 \cdot \frac{2}{\pi} \cdot \arctan 10t$.

To implement the DFNG model in the SPICE platform, the equations in the DFNG model need to be represented by physical equivalent circuit components. Specifically, the ferroelectric capacitor is modeled with a linear capacitor $C_{DE}$ and a switching element modeled as a voltage-controlled current source $I_{FE}$, as shown in **Figure S7a**. $I_{FE}$ is determined by a virtual voltage $V_f$, which is in turn modeled via a virtual integrator and two voltage-controlled voltage sources, as shown in **Figure S7b**. The integrator sub-circuit consists of a voltage-dependent current source $I_{int} = \exp\left(-\frac{V_a}{V_{FE}}\right)$ that feeds into a capacitor to continuously integrate the switching history based on the applied voltage $V_{FE}$, yielding an internal state variable $V_{out}$. Then, the first voltage-controlled voltage source converts $V_{out}$ into the variable $V_N = A^d {V_{out}}^d$, equivalent to the variable $\langle N(t) \rangle$ in the DFNG model. The second voltage-controlled voltage source generates $V_f = 1 - \exp(-V_N)$, which represents the fraction of switched region $f$ in the DFNG model. $V_f$ is then used to produce the ferroelectric switching current $I_{FE}$ in the sub-circuit in Figure S7a. Response from the linear capacitor $C_{DE}$ and $I_{FE}$ represents the dynamic FeCAP device, and is fed back to the main equivalent circuit to update the new $V_{FE}$ values.

The FeCAP cell SPICE model is then used to build circuit simulators in SPICE to evaluate 3×3 and 32×32 FeCap crossbar arrays. To accurately reflect physical routing within the array, interconnect parasitic components in the crossbar arrays are also carefully modeled in SPICE. Specifically, the Pi-network model is used to estimate the distributed resistance and capacitance between adjacent cross-point cells along both the word and bit lines shown in **Figure 6**. A standard $V_{DD}/2$ write scheme is implemented to assess the target cell's switching dynamics and evaluate disturbance on unselected cells. Operating at $V_{DD} = 1.5$ V, the selected word line is driven by the abovementioned arctangent-shaped pulse, while the selected bit line is grounded. Simultaneously, all unselected lines are biased at ½ $V_{DD} = 0.75$ V to suppress unintended polarization reversal.

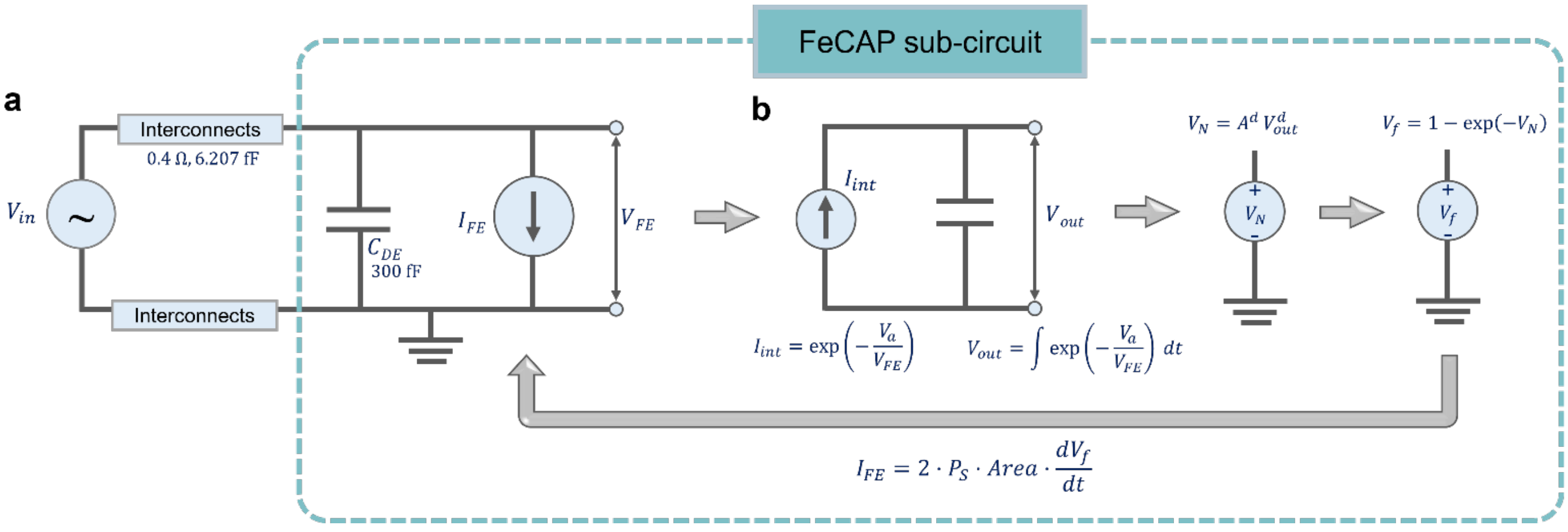

**Figure S7**: **Schematic of the SPICE implementation of the DFNG model to form several sub-circuits that represent a FeCAP cell.** The equations of the DFNG model equations are implemented using corresponding circuit components and sub-circuits as discussed in the text.

## 7. Fitting weight analysis

A weight of 2 has been added to the data points during 0-500 ps for the model fit for HZO and BTO. The primary motivation for introducing weight is to capture the polarization evolution during the rapid voltage variation, which occurs predominantly within the initial ~200 ps. This regime contains significantly fewer data points compared to the full transient, where the time to reach 95% switching ranges from approximately 1.3 ns to 9 ns across the measured cases. To account for ambiguity in defining the rise time of a distorted waveform, this weighted region is extended to 500 ps in the fitting procedure. Without weighting, the fitting algorithm tends to prioritize agreement in the post-500-ps regime.

Here we use the example of HZO to quantify the influence of weighting. The factor applied to the initial 500 ps is systematically varied, and both the total residual sum of squares (RSS) and the RSS in the 0-500 ps window are evaluated (**Figure S8**). As expected, increasing the weight improves agreement within the first 500 ps, but leads to a higher total RSS due to increased deviations towards the end of switching. A weighting factor of 2 is selected as a practical compromise, yielding a closer match in the early-time dynamics while maintaining acceptable overall agreement.

Despite this sensitivity in the fitting procedure, the key physical trends extracted from the model remain robust. Across all weighting conditions examined, the inferred growth dimensionality remains below 1, indicating strongly impeded domain growth. The activation voltage remains significantly higher than the applied voltage and <10% difference for low weights. The derived heterogeneous nucleation density is on the order of 0.2-1 $nm^{-d}$, which is within a reasonable fluctuation range.

Regarding the fit to AlBN, no weight is added to a specific time regime. The rise time (100 ns) is orders of magnitude shorter than the full switching time (10s μs) and thus the switching dynamic is almost unaffected by the rise time region. However, a weight of 10000 is added to all the normalized current data ($df/dt$), as the original values are 4 orders of magnitude smaller than the transformed fraction and the Avrami exponent. The addition of this strong weight is to ensure all the data points can contribute to the fit approximately to the same extent.

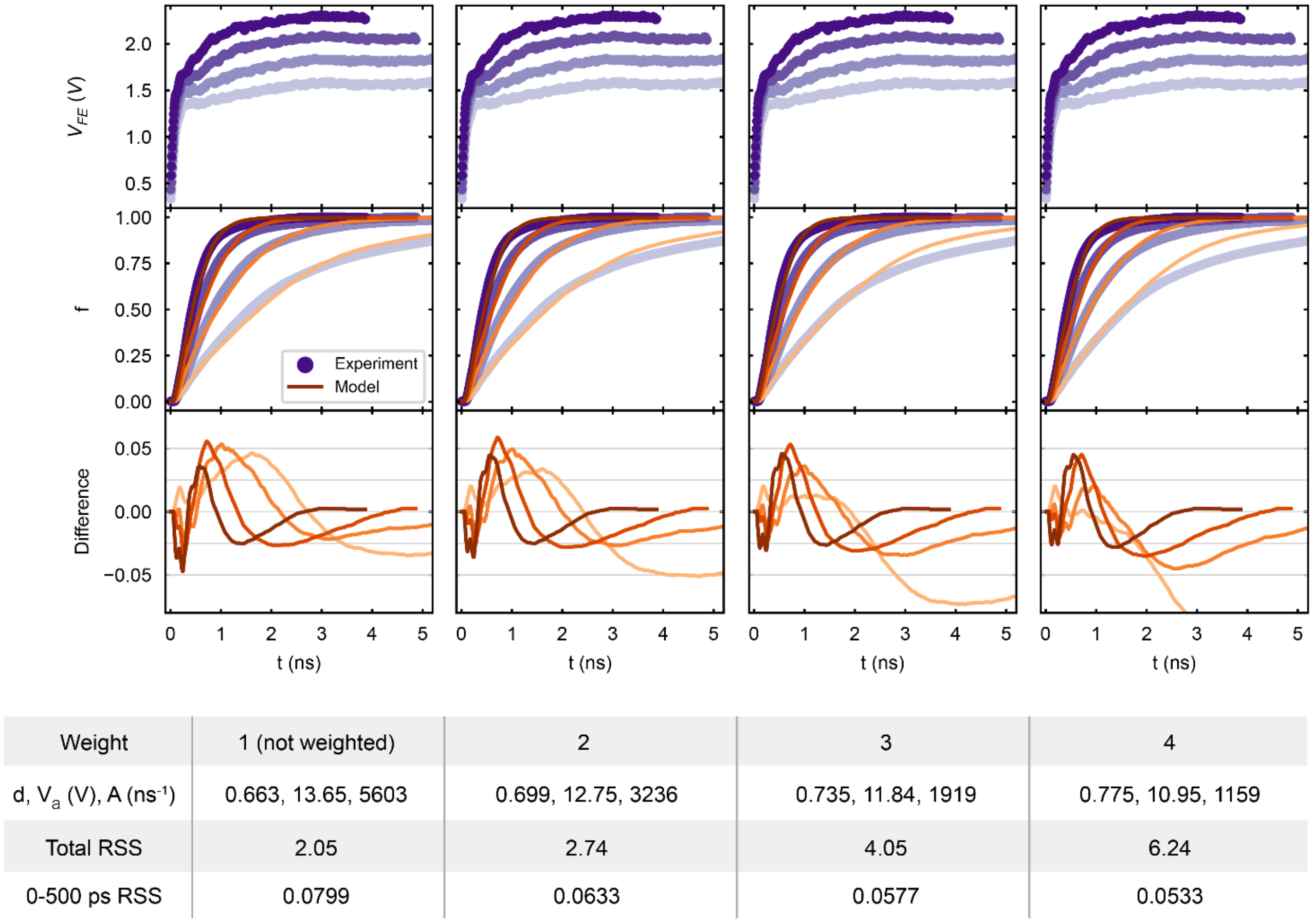

| Weight | 1 (not weighted) | 2 | 3 | 4 |
|---|---|---|---|---|
| d, $V_a$ (V), A ($ns^{-1}$) | 0.663, 13.65, 5603 | 0.699, 12.75, 3236 | 0.735, 11.84, 1919 | 0.775, 10.95, 1159 |
| Total RSS | 2.05 | 2.74 | 4.05 | 6.24 |
| 0-500 ps RSS | 0.0799 | 0.0633 | 0.0577 | 0.0533 |

**Figure S8: Analysis of fitting method.** The difference between the model-generated and experimentally measured transform fractions are plotted as a function of time, with varying weights from 1 to 4 added to the initial 500-ps stage. The associated fit parameters, total residual sum of squares (RSS) and RSS of the initial 500-ps regime are summarized in the table below.